\documentclass[slac_one]{revtex4}
\usepackage{graphicx}
\usepackage{fancyhdr}
\begin{document}

\title{{\small{2005 International Linear Collider Workshop - Stanford,
U.S.A.}}\\ 
\vspace{12pt}
Identification of Large Extra Spatial Dimensions at the LC}

%

\author{A. A. Pankov}
\affiliation{Pavel Sukhoi Technical University, Gomel 246746, Belarus}
\author{N. Paver}
\affiliation{University of Trieste and INFN, 34100 Trieste, Italy}

\begin{abstract}
We discuss the possibility to cleanly distinguish, in electron-positron
annihilation into fermion pairs at a high energy collider, the indirect
manifestations of graviton exchange from those of four-fermion contact
interactions. The method is based on cross section asymmetries emphasizing
the spin-2 character of graviton exchange and is explicitly applied to
the ADD scenario for gravity in extra dimensions. The availability of initial
beams longitudinal polarization is also taken into account in the analysis.
For typical c.m. energies and time-integrated luminosities foreseen at the
International Linear Collider ($\sqrt s=0.5-1\hskip 2pt{\rm TeV}$,
${\cal L}_{\rm int}=50-1000\hskip 2pt{\rm fb}^{-1}$), a 5$\sigma$
identification reach of $3.5-5.8\hskip 2pt{\rm TeV}$ for the mass scale $M_H$
relevant to the ADD model can be obtained, while the size of the reach on the
mass scales $\Lambda$ characterizing four-fermion contact interactions can be
of the order of  $45-65\hskip 2pt{\rm TeV}$.
\end{abstract}

\maketitle

\thispagestyle{fancy}


\section{INTRODUCTION}\label{sec:intro}
In a variety of proposed new physics (NP) scenarios, non-standard interactions
among the familiar Standard Model (SM) particles can be mediated by exchanges
of new quantum states with mass scales expected to be much larger than the
c.m. energy available at current (and perhaps future) colliders. Accordingly,
only indirect manifestations of such high mass scales and their
corresponding novel interactions can occur, through deviations of the measured
cross sections from the SM predictions. The most convenient theoretical
representation of such interactions is provided by the effective interaction
framework, where the non-standard Hamiltonian is expanded in a series of
specific local operators of increasing dimension, and accordingly the
transition amplitudes for processes among the SM particles are power
expanded in the (small) ratio between the `low' c.m. machine energy and the
relevant high mass scales. Generally, to limit the number of unknown
parameters to be constrained (or determined) experimentally, the
lowest-dimensional operator is retained in the expansion, assuming higher
powers to be negligible due to the strong suppression by the large mass scale.
Clearly, in this situation, NP searches are favoured by the signal enhancement
due to the high energies and luminosities available at the planned linear
colliders.

A relevant aspect in this regard is that, in principle, different kinds of
non-standard interactions may produce in the integrated cross sections similar
deviations from the SM and, therefore, it is important to devise suitable
observables that, given the expected experimental accuracy, can discriminate
among the various, and competing, possible sources of a given deviation. Here,
we will focus on the problem of cleanly identifying, in high energy
$e^+e^-\to f{\bar f}$ [$f\ne e,t$] at the International Linear Collider (ILC),
signals of the ADD model of gravity in large, compactified, extra spatial
dimensions with respect to the effects originating from  four-fermion
contact interactions. For this purpose, we shall use particular combinations
of integrated cross sections, the so-called ``center-edge asymmetries'',
sensitive to the angular dependence of deviations from graviton exchange.

We recall that the differential cross section of the considered processes
reads, in terms of helicity cross sections \cite{Schrempp}
($z\equiv\cos\theta$; $\alpha,\beta={\rm L,R}$):
\begin{equation}
\frac{d\sigma}{d z}=\frac{1}{4}\sum_{\alpha\beta}
\frac{d\sigma_{\alpha\beta}}{d z}; \qquad\qquad
\frac{d\sigma_{\alpha\beta}}{d z}=N_{\rm colors}\frac{3}{8}\sigma_{\rm pt}
\vert {\cal M}_{\alpha\beta}\vert^2\, (1\pm z)^2,
\label{crossdif}
\end{equation}
where `$\pm$' refer to the $\rm LL,RR$ and $\rm LR,RL$ configurations,
respectively. The helicity amplitudes can rather generally be expanded
into the familiar $\gamma,Z$ $s$-channel exchanges plus deviations induced
by the novel interaction ($\chi_Z(s)$ is here the $Z$ propagator):
\begin{equation}
{\cal M}_{\alpha\beta}\equiv{\cal M}_{\alpha\beta}^{\rm SM}
+\Delta_{\alpha\beta}= Q_e Q_f+g_\alpha^e\,g_\beta^f\,\chi_Z(s)+
\Delta_{\alpha\beta}.
\label{amplit}
\end{equation}
It may be useful to also reacall that the SM cross section can be decomposed
into $z$-even and $z$-odd parts through the total and forward-backward cross
sections as
\begin{equation}
\frac{d\sigma^{\rm SM}}{d z}=\frac{3}{8}\sigma^{\rm SM}
\left(1+z^2\right)+\sigma_{\rm FB}^{\rm SM}z.
\label{dsigmasm}
\end{equation}

In the ADD large extra dimension scenario \cite{Arkani-Hamed}, only gravity
can propagate in at least two extra spatial dimensions compactified to a
radius $R$ of the millimeter size or less, while the SM particles live in the
ordinary four-dimensional spacetime and their mutual gravitational
interactions are represented by the exchange of a tower of graviton
Kakuza-Klein (KK) states $\vec n$, very weakly [gravitationally] coupled and
with evenly spaced (and almost continuous) mas spectrum
$m_{\vec n}^2={\vec n}^2/R^2$ \cite{Giudice}. The summation over the KK
spectrum requires the introduction of a ultaviolet cut-off mass scale $M_H$,
expected in the (multi) TeV region, and the interaction can be represented by
a dimension-8 effective Lagrangian of the form
\cite{Hewett}
\begin{equation}
{\cal L}^{\rm ADD}=i\frac{4\lambda}{M_H^4}T^{\mu\nu}T_{\mu\nu},
\label{dim-8}
\end{equation}
where $T_{\mu\nu}$ is the energy-momentum tensor and
$\lambda=\pm1$.\footnote{In principle, a smooth cutoff procedure based on
the ``minimal length scale'' can be applied in the sum over KK states
\cite{Hossenfelder}.}
The corresponding deviations in Eq.~(\ref{amplit}) are $z$-dependent:
\begin{equation}
\Delta_{\rm LL}({\rm ADD})=\Delta_{\rm RR}({\rm ADD})=f_G(1-2z),\quad
\Delta_{\rm LR}({\rm ADD})=\Delta_{\rm RL}({\rm ADD})=-f_G(1+2z),
\label{ADD}
\end{equation}
where $f_G=\lambda\,s^2/(4\pi\alpha_{\rm e.m.}M_H^4)$ represents the strength
of the interaction associated with spin-2 graviton exchange.

The four-fermion contact interaction scenario (CI) can be represented by the
following vector-vector dimension-6 effective Lagrangian, and corresponding
helicity amplitudes deviations from the SM
($\vert\eta_{\alpha\beta}\vert=1,0$) \cite{Eichten}:
\begin{equation}
{\cal L}_{\rm CI}
=4\pi \sum_{\alpha,\beta}\hskip 2pt
\frac{\eta_{\alpha\beta}}{\Lambda^2_{\alpha\beta}}
\left(\bar e_\alpha\gamma_\mu e_\alpha\right)
\left(\bar f_\beta\gamma^\mu f_\beta\right);\qquad
\Delta_{\alpha\beta}({\rm CI})=\pm\frac{s}{\alpha_{\rm e.m.}}
\frac{1}{\Lambda^2_{\alpha\beta}}.
\label{CI}
\end{equation}
As one can see, in this case amplitudes deviations are $z$-independent.
Actually, although originally inspired by fermion compositeness remnant
binding forces, ${\cal L}_{\rm CI}$ should more generally be considered as an
effective, ``low energy'' representation of a variety of non-standard
interactions acting at energy scales $\Lambda$ much larger than the process
Mandelstam variables, for example the exchanges of very heavy $Z^\prime$s
\cite{Leike}, leptoquarks \cite{Riemann} and even scalar particle exchanges
in the $t$-channel, such as sneutrinos \cite{Rizzo} in the contact interaction
limit.

Clearly, suitable observables are needed to discriminate signals of the
different kinds of NP models.

\section{CENTER-EDGE ASYMMETRY}\label{sec:ace}
We consider the difference between the ``central'' and ``edge'' parts of the
cross section:
\begin{equation}
\sigma_{\rm CE}(z^*)\equiv\sigma_{\rm C}-\sigma_{\rm E}
=\left[\int_{-z^*}^{z^*}-
\left(\int_{-1}^{-z^*}+\int_{z^*}^{1}\right)\right]
\frac{d\sigma}{d z}\,{d z},
\label{sce}
\end{equation}
with $0<z^*<1$, and define the asymmetry $A_{\rm CE}$ by the ratio
\cite{Osland}
\begin{equation}
A_{\rm CE}=\frac{\sigma_{\rm CE}}{\sigma}.
\label{ace}
\end{equation}
This asymmetry is sensitive only to the $z=\cos\theta$-even terms of the
differential cross section. Indeed, Eq.~(\ref{CI}) shows that in the
four-fermion interaction case the differential cross section will have exactly
the same angular dpendence as the SM one [see Eq.~(\ref{dsigmasm})],
therefore:
\begin{equation}
A_{\rm CE}^{\rm CI}(z^*)=A_{\rm CE}^{\rm SM}=\frac{1}{2}\,z^*\,({z^*}^2+3)-1,
\label{ACECI}
\end{equation}
independent of $\sqrt s$, flavour of final fermions and longitudinal
polarization. Accordingly, as regards the deviation from the SM prediction,
$\Delta A_{\rm CE}\equiv A_{\rm CE}-A_{\rm CE}^{\rm SM}$, for any value of
$z^*$:
\begin{equation}
\Delta A_{\rm CE}^{\rm CI}\equiv A_{\rm CE}^{\rm CI}-A_{\rm CE}^{\rm SM}=0.
\label{deviat-ace-CI}
\end{equation}
Also, from Eq.~(\ref{ACECI}) one notices that for the particular value
$z^*_0=(\sqrt{2}+1)^{1/3}-(\sqrt{2}-1)^{1/3}\simeq 0.60$
($\theta\simeq 53.4^\circ$), one has
$A_{\rm CE}^{\rm SM}(z^*_0)=A_{\rm CE}^{\rm CI}(z^*_0)=0$.

By contrast, Eq.~(\ref{ADD}) implies that finite deviations of $A_{\rm CE}$
occur for the graviton-exchange ADD scenario for all $z^*$, and these are
dependent on the flavour of the final states $f$. Indeed, at the leading
order in the graviton coupling $f_G$, i.e., retaining in the differential
cross section interference terms with the SM only:
\begin{equation}
\Delta A_{\rm CE}^{\rm ADD}(z^*)\equiv A_{\rm CE}^{\rm ADD}-A_{\rm CE}^{\rm SM}
\propto f_Gz^*\left(1-z^{*2}\right).
\label{deladd}
\end{equation}

In conclusion, the asymmetry $A_{\rm CE}$ is ``blind'' to conventional
four-fermion contact interactions at all $z^*$ (no deviation from the SM in
this case), whereas it is sensitive to ADD graviton exchange effects and,
accordingly, can cleanly identify the new physics represented by this
scenario. Moreover, the maximal senstitivty could be obtained by measuring
$A_{\rm CE}$ around $z^*_0$ where the SM and the CI contributions are both
vanishing.\footnote{This observable can similarly be applied to identify
graviton exchange in lepton-pair production at hadron
colliders \cite{Dvergsnes}.}

\section{CENTER-EDGE FORWARD-BACKWARD ASYMMETRY}\label{sec:acefb}
With $0<z^*<1$, we now consider the analogue of Eq.~(\ref{sce})
\begin{equation}
\sigma_{\rm CE,FB}\equiv\left(\sigma_{\rm C,FB}-\sigma_{\rm E,FB}\right)
=\left[\left(\int_0^{z^*}-\int_{-z^*}^0\right) \,
- \, \left(\int_{z^*}^1-\int_{-1}^{-z^*}\right)\right]\hskip 2pt
\frac{d\sigma}{d z}\hskip 2pt {d z}.
\label{scefb}
\end{equation}
This asymmetry is defined by the ratio, sensitive to $z=\cos\theta$-odd terms
only \cite{Pankov}:
\begin{equation}
A_{\rm CE,FB}=\frac{\sigma_{\rm CE,FB}}{\sigma}.
\label{acefb}
\end{equation}

For the case of four-fermion contact interactions, Eq.~(\ref{CI}), due to the
identical angular dependence as in the SM, one immediately finds the relations
\begin{equation}
A_{\rm CE,FB}^{\rm SM}(z^*)=A_{\rm FB}^{\rm SM}\,(-1+2{z^*}^2) \qquad
\quad \Longrightarrow\qquad\quad
A_{\rm CE,FB}^{\rm CI}(z^*)= A_{\rm FB}^{\rm CI}\,(-1+2{z^*}^2).
\label{acefbsm}
\end{equation}
Correspondingly, in general contact interactions determine finite deviations
of $A_{CE,FB}^{\rm CI}$ from the SM predictions, as indicated by
Eq.~(\ref{acefbsm}). However, at the value
$z^*_{\rm CI}=1/\sqrt 2$ $(\theta=45^\circ)$ one has
$A_{\rm CE,FB}^{\rm SM}(z^*_{\rm CI})=A_{\rm CE,FB}^{\rm
CI}(z^*_{\rm CI})= \Delta A_{\rm CE,FB}^{\rm CI}(z^*_{\rm CI})=0$,
i.e., no deviation there. Consequently, $A_{\rm CE,FB}\ne 0$ at this
value of $z^*$ would definitely signal the presence of NP different from
four-fermion contact interactions.

\begin{figure*}[t]
\centering
\includegraphics[width=100mm]{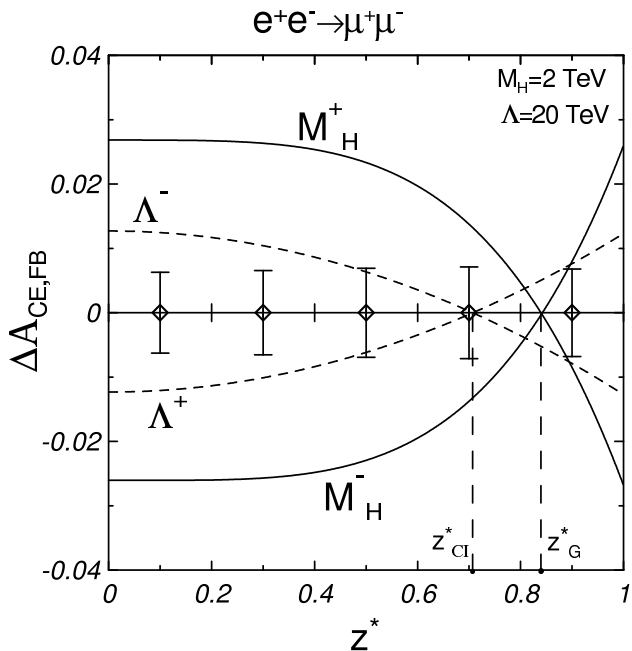}
\vspace{-4.5cm} \caption{$\Delta A_{\rm CE,FB}(z^*)$ in the CI and
the ADD scenarios for the values of $\Lambda$ and $M_H$ indicated in the text.
The $\pm$ superscripts indicate positive and negative interference with
the SM, respectively. The vertical bars are the statistical uncertainty at a
ILC with $\sqrt s=0.5$ TeV and ${\cal
L}_{\rm int}=50\, fb^{-1}$. }
\label{Delta-acefb}
\end{figure*}

In the graviton exchange ADD model, using Eq.~(\ref{ADD}) one directly finds
for the deviation of $A_{\rm CE,FB}(z^*)$ from the SM the following expression
to leading order in $f_G$:
\begin{equation}
\Delta A_{\rm CE,FB}^{\rm ADD}(z^*)= \Delta A_{\rm FB}^{\rm ADD}
\left(-1+2{z^*}^4\right)\qquad
\left[\Delta A_{\rm FB}^{\rm ADD}\propto f_G\right].
\label{deviat-acefb-add}
\end{equation}
Eq.~(\ref{deviat-acefb-add}) shows that $\Delta A_{\rm CE,FB}^{\rm
ADD}(z^*_{\rm G})=0$ for $z^*_{\rm G}=2^{-1/4} \simeq 0.84$
($\theta\simeq 33^\circ$), i.e., no deviation from graviton
exchange at that value of $z^*$ but possible deviations from contact
interactions can occur there. The behaviour with $z^*$ of
$\Delta A_{\rm CE,FB}(z^*)$ in the two NP scenarios considered here is
shown in the case of annihilation into muon pairs in Fig.~\ref{Delta-acefb},
for $M_H=2\hskip 2pt {\rm TeV}$ and $\Lambda=20\hskip 2pt{\rm TeV}$ as
illustrative examples. The `+' and `-' superscripts indicate positive or 
negative interference with the SM. To show their statistical significance, 
deviations are compared to the statistical uncertainty expected at the ILC with
$\sqrt s=0.5\hskip 2pt{\rm TeV}$ and
${\cal L}_{\rm int}=50\hskip 2pt{\rm fb}^{-1}$, that are typical planned
values \cite{Abe}.
 \par
The above considerations, and Fig.~\ref{Delta-acefb}, then suggest
the following kind of analysis. The measurement of $A_{\rm CE,FB}$
at $z^*\approx z^*_{\rm CI}$ or below has maximal sensitivity to
the ADD graviton exchange scenario, with no (or minimal)
contamination from CI; this measurement can be combined with the
measurement of $A_{\rm CE}$ to further enhance the identification
reach on $M_H$. Instead, the measurement of $A_{\rm CE,FB}$ in an
interval around $z^*\approx z^*_{\rm G}$ should have maximal
sensitivity to four-fermion CI, hence to the scales $\Lambda$,
with least contamination from ADD effects. Inclusion of beams longitudinal
polarization is easily obtained and results, for our basic
observables $A_{\rm CE}$ and $A_{\rm CE,FB}$, into the same $z^*$
dependence as found above times a factor accounting for the
initial spin configurations \cite{Pankov}.

\section{IDENTIFICATION REACHES ON THE MASS SCALES}\label{sec:reaches}

Basically, in order to evaluate the potential identification reach on the 
fundamental mass parameters $M_H$ and $\Lambda$ at the ILC, as achieved by
measurements of the asymmetries defined above, one should compare the
deviations from the SM predictions to the expected experimental uncertainties
on those observables. One can apply a conventional $\chi^2$ analysis, where
the $\chi^2$ can be formally defined as
\begin{equation}
\chi^2=\frac{\left(\Delta{\cal O}^f\right)^2}{\left(\delta{\cal O}^f\right)^2},
\label{chisquared}
\end{equation}
with ${\cal O}=A_{\rm CE}, A_{\rm CE,FB}$, $\Delta{\cal O}^f$ are the
deviations of the asymmetries previously discussed, and $\delta{\cal O}^f$
are the experimental uncertainties. In practice, several final states $f$ and
$A_{\rm CE}$ with $A_{\rm CE,FB}$ themselves can be appropriately combined in
Eq.~(\ref{chisquared}). Constraints on $M_H$ and $\Lambda$s will then follow
from the condition $\chi^2\le\chi^2_{\rm C.L.}$, where the actual value of
$\chi^2_{\rm C.L.}$ depends on the desired confidence level.

We will here consider an ILC with c.m. energy of either 0.5 Tev or 1 TeV and in
both cases electron and positron longitudinal polarizations
$\vert P_1\vert=0.80$ and $\vert P_2\vert = 0.60$, and will plot the
numerical results on $M_H$ and $\Lambda$ for time-integrated luminosity in
the range $50-1000$ ${\rm fb}^{-1}$. Regarding systematic uncertainties, one
expects them to largely cancel in the ratios (\ref{ace}) and (\ref{acefb}), 
and thus the statistical uncertainties to dominate the experimental 
uncertainty. Indeed, the major sources of systematic uncertainties are found 
to originate from the errors in the luminosity and in the degree of initial 
beams longitudinal polarization, for which we assume
$\Delta{\cal L}_{\rm int}/{\cal L}_{\rm int}=\Delta P_1/P_1=\Delta P_2/P_2=
0.5$\% (more details can be found in \cite{Pankov}).

\begin{figure*}[t]
\centering
\includegraphics[width=80mm]{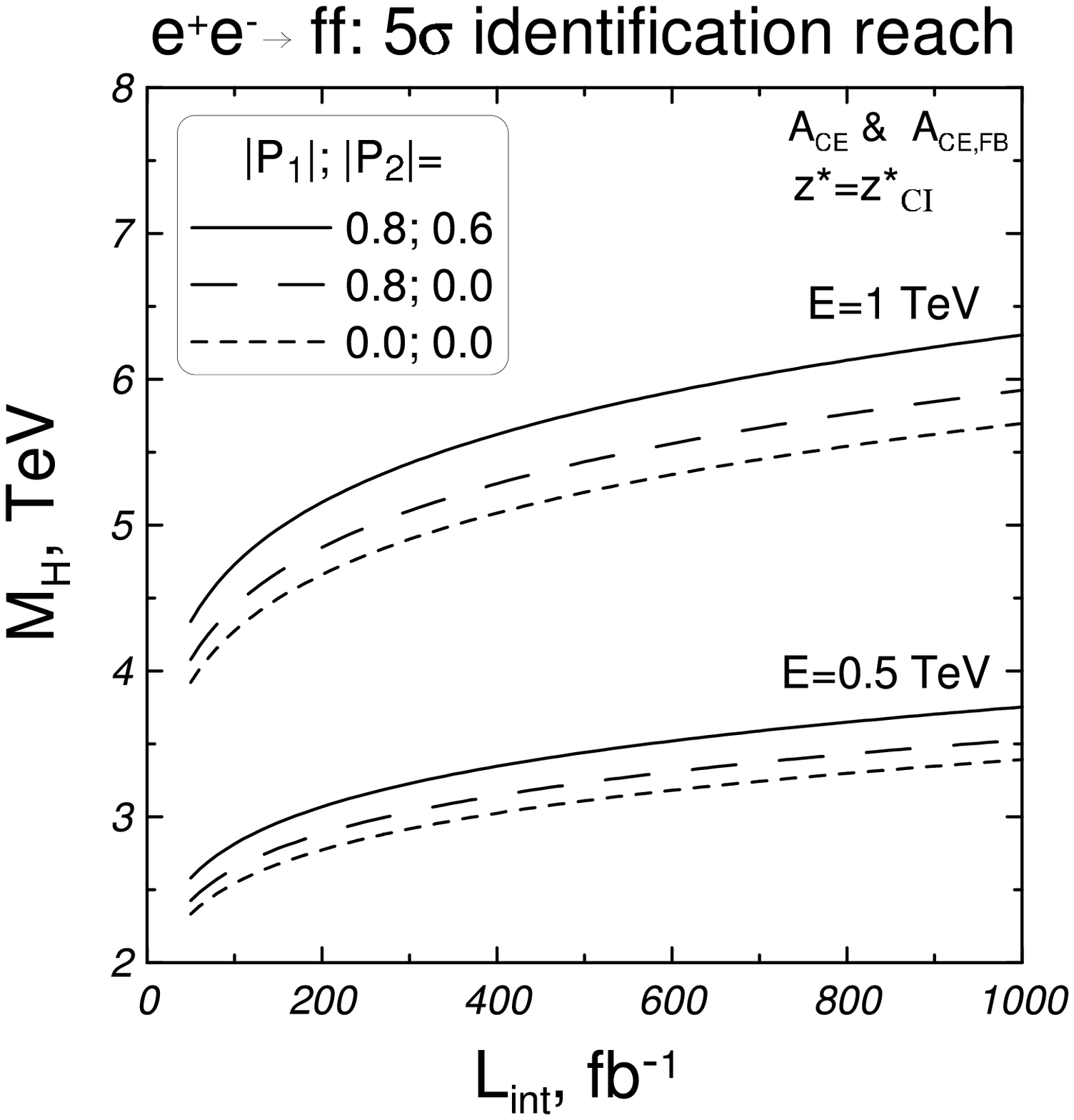}
\includegraphics[width=80mm]{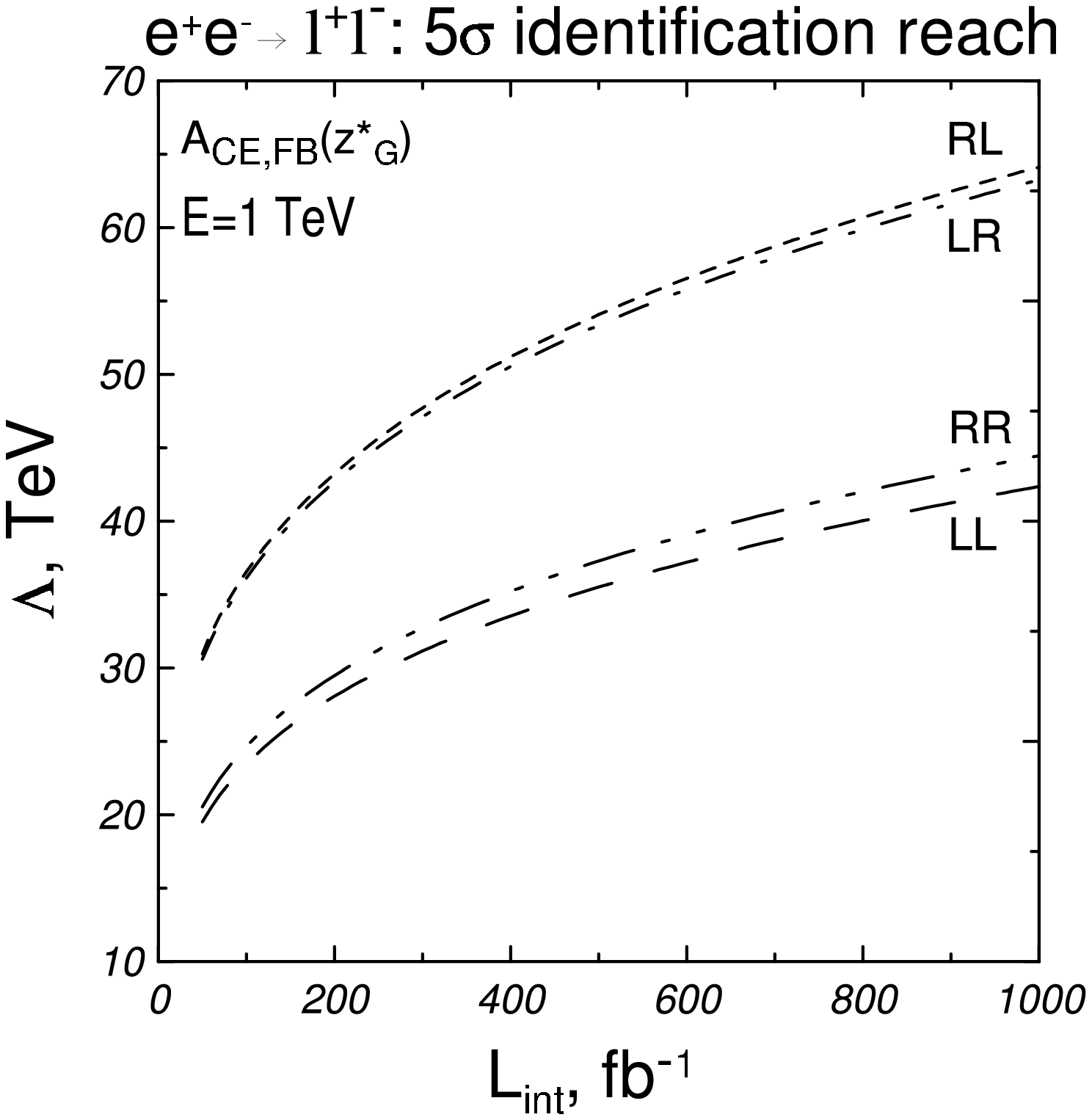}
\vspace{-2.5cm} \caption{Left panel: 5$\sigma$ identification
reach on the mass scale $M_H$ {\it vs.}\ integrated luminosity.
Right panel: 5$\sigma$ reach on the mass scales
$\Lambda_{\alpha\beta}$ {\it vs.}\ integrated luminosity, labels
attached to the curves indicate the helicity configurations
$\alpha\beta={\rm LL,RR,LR,RL}$, see Eq.~(\ref{CI}). } \label{ID}
\end{figure*}

In Fig.~\ref{ID} [left panel], the 5$\sigma$ identification reach on the mass
scale $M_H$ relevant to graviton exchange is shown as a function of the
luminosity, and for different longitudinal beam polarization configurations.
Here, the final annihilation $f\bar f$ channels with $f=\mu,\tau, b,c$ have
been summed over, and the asymmetries $A_{\rm CE}(z^*_{\rm CI})$ and
$A_{\rm CE,FB}(z^*_{\rm CI})$ have been combined. As pointed out in previous
sections, this provides the maximal sensitivity to the ADD graviton exchange
scenario, with no contamination from four-fermion contact interactions.
One can see that the identification reach on $M_H$ at the 5$\sigma$ level
is of the order of $3.5-5.8$ TeV for energies between 0.5 TeV and 1 TeV and a
luminosity of about 500 ${\rm fb}^{-1}$, and can potentially increase to
$(6.3-7.5)\times E_{\rm c.m.}$ for the highest luminosity. This should be 
compared with the current limit from LEP and Tevatron,
$M_H\ge 1.10-1.28$ TeV \cite{Ask}. Also, one can notice the (slow) scaling of
$M_H$ $\sim\left(s^3{\cal L}_{\rm int}\right)^{1/8}$, reflecting the (high)
dimension of the effective interaction of Eq.~(\ref{dim-8}).

In the right panel of Fig.~\ref{ID}, we show the 5$\sigma$ reach on the
four-fermion interaction mass scales $\Lambda$ as a function of luminosity,
at the ILC c.m. energy of 1 TeV. Here, the observable
$A_{\rm CE,FB}(z^*_{\rm G})$ is used, and only the final $l^+l^-$
pairs with $l=\mu,\tau$ are combined in the $\chi^2$. Also, the longitudinal
polarizations are chosen as $P_1=0.80$ and $P_2=-0.60$, to disentangle the
various helicity combinations of Eq.(\ref{CI}). It can be seen that the limits
on $\Lambda$s scale as $\sim\left(s{\cal L}_{\rm int}\right)^{1/4}$, faster
than for $M_H$, due to the (lower) dimension-6 of the effective interaction
(\ref{CI}). According to the previous discussion, maximal sensitivity to
four-fermion CI, with least (or no) contamination from ADD graviton exchange
is expected. The potential 5$\sigma$ reach on $\Lambda$s of the linear
collider ranges up to 45 TeV and 65 TeV for c.m. energies of 0.5 TeV
and 1 Tev, respectively, depending on the particular helicity configurations.
Current bounds, of the order of 10 TeV and depending on the CI model 
considered, are reviewed in \cite{Eidelman}. Also, the limits on $\Lambda$ 
obtained here may potentially improve the constraints on a very heavy 
sneutrino parameters \cite{Pankov}.

\begin{acknowledgments}
The work of NP was supported by funds of the University of Trieste and of the
MIUR (Italian Ministry for University and Research).
\end{acknowledgments}


\end{document}